# Interface-dependent Phase Transitions and Ultrafast Hydrogen Superionic Diffusion of $H_2O$ Ice

Pengfei Hou[1], Yumiao Tian[1], Zifeng Liu[1], Junwen Duan[1], Hanyu Liu[1,4,*], Xing Meng[1,2,†], Russell J. Hemley[3], Yanming Ma[1,4,5,‡]

[1]*Key Laboratory of Material Simulation Methods and Software of Ministry of Education, State Key Laboratory of High Pressure and Superhard Materials, College of Physics, Jilin University, Changchun 130012, China.*
[2]*Key Laboratory of Physics and Technology for Advanced Batteries (Ministry of Education), College of Physics, Jilin University, Changchun 130012, China.*
[3]*Department of Physics, Department of Chemistry, and Department of Earth and Environmental Sciences, University of Illinois Chicago, Chicago, IL 60607.*
[4]*International Center of Future Science, Jilin University, Changchun 130012, China.*
[5]*School of Physics, Zhejiang University, Hangzhou, China.*

**ABSTRACT**. High-pressure experiments using diamond anvils have revealed novel properties and phase behavior of $H_2O$ under extreme conditions. When contained in diamond-anvil cells, the $H_2O$ samples are usually in direct contact with the diamond anvil. However, the extent to which this interface affects measured pressure-induced properties and behavior, including coexistence lines of ice phases, remains unknown. Combining artificial neural network methods and active learning schemes with large-scale molecular dynamics simulations, we elucidate the interfacial effects on various properties of high-pressure ice phases, including superionic states, solid-solid phase transitions, and melting. The results reveal that the presence of this interface can significantly lower the hydrogen superionic transition temperature. Remarkably, the interface can also induce a spontaneous transition from bcc- to fcc-based ice following the inverse Bain mechanism. Further, we redefined a stability field of bcc and fcc ice below the melting line and predicted the existence of fcc ice at much lower pressures than previously thought. More broadly, the results emphasize the importance of interface effects in understanding a wide range of phenomena reported in experimental studies of ice under pressure, including inconsistencies between theoretical and experimental results of this fundamental system.

## I. INTRODUCTION.

The novel properties and behavior of $H_2O$ under extreme conditions have attracted intensive attention due to the pivotal role that this ubiquitous material plays in Nature [1-4]. As a prevalent substance, $H_2O$ exhibits extraordinarily rich polymorphism containing more than 20 experimentally confirmed liquid and solid phases, the latter characterized by different oxygen sublattices and hydrogen distributions within these sublattices [5-9]. Intriguingly, $H_2O$ ices above 3 GPa have oxygen sublattices that are body-centered cubic (bcc) arrays (or slight distortions thereof) [10] except for a recently discovered cubic superionic ice with a face-centered cubic (fcc) sublattice [11-13].

Static compression techniques have been key to discovering the polymorphism, phase behavior, and other properties of $H_2O$ ice under extreme conditions [5,12,13]. Meanwhile, theoretical calculations have also been devoted to disclosing the mysteries of ice and exhibited great potential in exploring the novel phenomena, including VII-VII´-X transition, VII´–VII´´ transition, and melting relations [14-16]. However, until now there remain many mysteries among the high-pressure ice phases, particularly regarding the elusive location of various coexistence lines of different forms of ice. For example, significant uncertainties persist regarding the melting curve of ice: reported melting temperatures at 20 GPa vary by up to 300 K, with this discrepancy increasing to 800 K at 40 GPa [17-27]. The bcc-fcc ice phase transition provides another interesting case: theoretical predictions suggest fcc ice stability to pressures above 100 GPa [28,29], yet experimental evidence shows this phase can form at much lower pressures - with some studies reporting detection of the phase as low as 45 GPa or even 29 GPa [12,13,27]. The discrepancies in reported phase coexistence lines, both across experimental studies and between theoretical and experimental results, highlight the need for more definitive studies.

In high-pressure diamond-anvil cell (DAC) experiments on $H_2O$ ice, the sample is typically in direct contact with the diamond, usually the (100) lattice plane. Rare exceptions are experiments in which a separate pressure-transmitting medium is used, though many of the standard quasi-hydrostatic media react with the $H_2O$ (*e.g.*, to form dense clathrates [30]). As a result, experiments at the highest pressure have been conducted without a pressure-transmitting medium. Studies from other fields [31,32] have shown that water molecules in the contact interface exhibit distinct structural arrangements and physicochemical properties that differ significantly from those in the bulk water region. Considering the extreme conditions usually ranging from tens to even hundreds of gigapascals, water molecules at the interface may react with diamond anvil to form interfacial compounds because of the pressure-induced changes in chemical reactivity [33]. In fact, as early as 1989, Hemley et al. discovered that the presence of the interface could also promote the crystallization of high-density amorphous ice to ice VII', which occurs at ~4 GPa (at 77 K), followed by the VII'-VIII transition on warming at these low pressures. [4] Whether this interface will further affect the phase transition behavior and measurements of various coexistence lines of the ice phase remains unknown.

[*]Contact author: hanyuliu@jlu.edu.cn
[†]Contact author: mengxing@jlu.edu.cn
[‡]Contact author: mym@jlu.edu.cn

Here we constructed a neural network potential (NNP) for diamond, bcc ice, and their prevalent interface models diamond (100) /bcc ice (100) over a wide P-T range using artificial neural networks and active learning scheme. Utilizing large-scale molecular dynamics (MD) simulations, we revealed the interfacial dependence of various properties of bcc ice including superionic states, solid-solid phase transitions, and melting.

## II. RESULTS

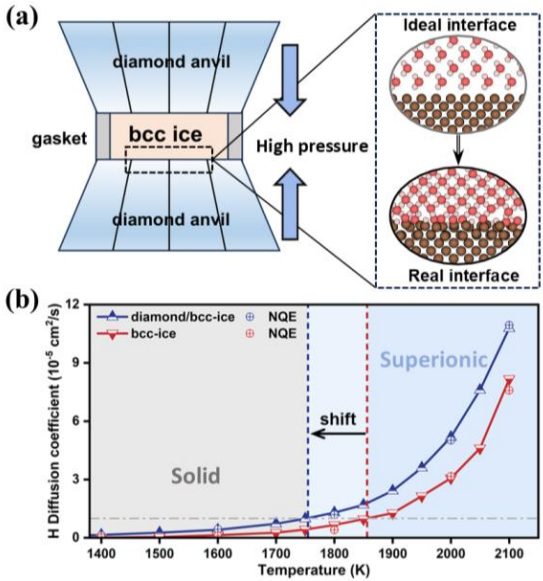

FIG. 1. Interface reconstruction and ultrafast hydrogen superionic diffusion. (a) Schematic diagram of interface reconstruction between diamond and bcc ice under extreme conditions. (b) Hydrogen diffusion coefficients for the two models - bcc ice and diamond (100) /bcc ice (100) interface - with and without nuclear quantum effects from 1400 K to 2100 K.

Fig. 1(a) illustrates such generally overlooked reconstruction of the interface between diamond and ice when compressed. Here we primarily selected the common bcc-based phase ice VII and constructed an interface model - diamond (100) /bcc ice (100), as presented in Fig. S1 [34]. The results of the geometry optimization and ab initio MD simulations revealed that water molecules close to the surface dissociated and hydrogen or oxygen bonded with carbon under high pressure but maintained the typical bcc oxygen sublattice. Despite the fact that this interface model is up to 624 atoms, the ice thickness is only 13 angstroms at 100 GPa, and it is difficult to accurately distinguish between the bulk and the interface region, which limits the fine study of interface effects on the various properties of ice.

Then we constructed a NNP for diamond, bcc ice, and their prevalent interface models diamond (100) /bcc ice (100) over a wide $P$-$T$ range using active learning scheme. Eventually, a high-quality NNP was obtained using 17076 pieces of DFT data collected from 66.41 million snapshots after 13 iterations of active learning. The detailed training process, NNP accuracy testing, and parameter settings are shown in the Supplemental Material [34]. Then, we constructed a large-scale bcc ice model (2000 $H_2O$ molecules, 6000 atoms total) and a diamond (100)/bcc ice (100) interface model (2420 C atoms, 6076 $H_2O$ molecules, 20,648 atoms total), as shown in Fig. S6, hereafter referred to as bcc-ice and diamond/bcc-ice, respectively. Especially such thick bcc ice in the interface model is sufficient to explore the interface effects on the various physical properties of ice.

We utilized the trained NNP to perform large-scale NPT MD simulations at 100 GPa and varying temperatures on the bcc-ice and diamond/bcc-ice systems. Fig. S7 displays the hydrogen mean square displacement (MSD) over time for both systems at various temperatures. The MSD of the diamond/bcc-ice system consistently exceeds that of bcc-ice at identical temperatures. Using the Einstein relation, the hydrogen diffusion coefficients are shown in Fig. 1(b). Obviously, the presence of the interface does significantly promote the hydrogen diffusivity and reduces the superionic transition temperature by about 104 K (take $1\times10^{-5}$ $cm^2$/s as the threshold). After accounting for nuclear quantum effects (NQEs), the diffusion coefficient remains almost unchanged, as indicated in Fig. 1(b) and Fig. S8, since the quantum effect is negligible compared to the excessive thermal effect of high temperatures. Furthermore, we performed the relevant MD calculations for the remaining pressures and temperature and obtained the superionic transition boundary (Fig. S9) of these two systems with varying pressures from 20 GPa to 120 GPa. The results revealed this interfacial effect was prevalent, with temperature reductions ranging from 44 K at 20 GPa to 118 K at 120 GPa. Notably, when taking a small threshold of $1\times10^{-6}$ $cm^2$/s, the temperature reduction reached 242 K at 120 GPa.

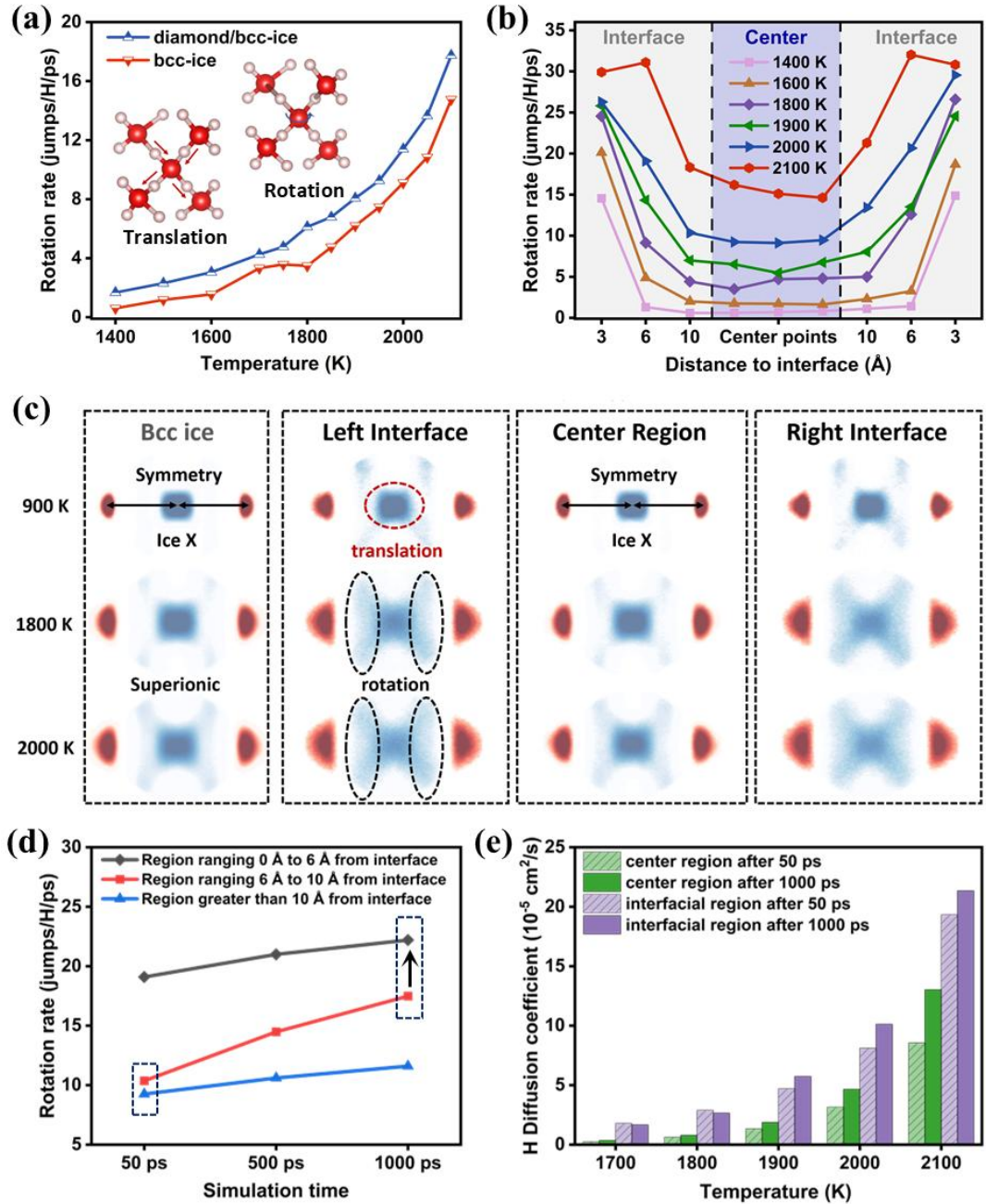

FIG. 2. Origin of ultrafast hydrogen diffusivity in the interface system. (a) The hydrogen rotation rate of bcc-ice and diamond/bcc-ice systems versus temperature. (b) Hydrogen rotation rate in different regions of diamond/bcc-ice system. (c) The projections of oxygen and hydrogen positions along O-H⋯O triplets at 100 GPa in bcc-ice and diamond/bcc-ice systems. (d) The variation of rotation rate versus simulation time in different regions of the diamond/bcc-ice system. (e) Hydrogen diffusion coefficient for the interfacial region within 15 Å from the interface and the other center region in diamond/bcc-ice system.

Hydrogen diffusion in ice can be viewed in terms of atom translation and molecular rotation. Translation is that hydrogen atoms jump between two O atoms from $O_a$-H⋯$O_b$ to $O_a$⋯H-$O_b$, but this $O_aHO_b$ triplet remains no change. While rotation stands for the situations in which the triplet will change from $O_a$-H⋯$O_b$ to $O_a$-H⋯$O_c$ along with rotation of hydrogen atoms. As shown in Fig. 2(a), the molecular rotation rate could serve as an excellent descriptor for the hydrogen diffusion coefficient because only by starting to rotate can hydrogen atoms diffuse from their initial positions and become delocalized. Hence, we present the hydrogen rotation rate in different regions of the diamond/bcc-ice system in Fig. 2(b), and the results show that the hydrogen rotation rate at the interface is significantly higher than that in the center region and decreases significantly with the increasing distance from the interface, but this decay rate will slow down with the increasing temperature. Fig. 2(c) shows the projections of oxygen and hydrogen positions along O-H···O triplets at 100 GPa in bcc-ice and diamond/bcc-ice systems, where the red and blue regions represent the probability density of oxygen and hydrogen atoms, respectively. Both systems exhibit the characteristic of the ice X phase at 100 GPa with symmetric hydrogen bonding. But for the diamond/bcc-ice system, the distribution of hydrogen at the interface region shows higher degree of delocalization caused by the decomposition of water molecules and the bonding with the C atoms near the diamond surface under high pressure.

The hydrogen diffusion of the interface system also exhibits more time-dependent characteristics. As shown in Fig. S10, for the diamond/bcc-ice system, the diffusion coefficients of 1000 ps MD simulation are appreciably larger than that of 50 ps MD simulation. Fig. 2(d) shows the variation of hydrogen rotation rate with simulation time in different regions of the diamond/bcc-ice system. The molecular rotation rates of all regions increase with the simulation time. The increasing slope of the region ranging 6 Å to 10 Å away from the interface [red solid line in Fig. 2(d)] is larger than the region ranging 0 Å to 6 Å [black solid line in Fig. 2(d)] or greater 10 Å [blue solid line in Fig. 2(d)]. The molecular rotation rates of the region ranging from 6 Å to 10 Å at 50 ps are close to that of the region greater 10 Å. In comparison, it rises and gradually approaches that of the region ranging 0 Å to 6 Å at 1000 ps [dotted-line rectangle labeled in Fig. 2(d)], which further revealed that hydrogen superionic diffusion of the region far away from the interface is gradually activated and gradually spread towards the center region. Subsequently, the contributions of the total hydrogen diffusion coefficient are also divided into the interfacial region and center region in Fig. 2(e), which indicates that the hydrogen diffusivity in the center region does increase with the simulation time and interface effect does affect the superionic diffusion behaviors of the center region farther from the interface. This also indicates that the interface effect on the properties of ice is by no means limited to the interface region, but may have more significant long-range effects on the properties of the whole ice.

The interface effect can be further extended to the observation of melting lines of ice. Fig. 3(a) displays the oxygen diffusion coefficient as a function of temperature at 20 GPa. MD simulations on homogeneous bcc-ice with a perfect lattice exhibit significant superheating effects [35,36] such that oxygen atoms still do not diffuse at 1200 K. However, oxygen atoms diffuse at about 900 K after introducing the interface. Its melting point is about 962 K (taking $10^{-5}$ cm$^2$/s as the melting threshold), which is in good agreement with experimental results (937 K at 20 GPa and 985 K at 21 GPa) [20]. This phenomenon shows that the disorder effect caused by the real diamond/ice interface in high-pressure experiments can effectively overcome the superheating effects in MD simulations, offering a feasible approach for the accurate estimation of melting points under high pressure.

More surprisingly, mixed phase bcc ice and fcc ice appear between 700 K and 900 K, as shown in Fig. 3(a). And Fig. 3(b) depicts the bcc to fcc transformation process in the diamond/bcc-ice system at 20 GPa and 900 K. At 10 ps, some fcc ice phases begin to appear in the center region and takes up 17 % of the simulation cell. As simulation time increases, the proportion of fcc ice also gradually rises, reaching 51% and basically no bcc ice phase at 50 ps. In subsequent 100 ps MD simulations (Fig. S11), the fcc ice region and the disordered ice region evolves into each other, which indicates that the proportion of fcc ice may be underestimated. The gray disordered regions are virtually caused by the oxygen atoms deviating from the equilibrium position driven by thermal effects but it can return to equilibrium fcc positions. Fig. S12 illustrates this transition of fcc to bcc on cooling and thus the reversibility of the process. Moreover, the complete phase transition from bcc ice to fcc ice also occurs at 850 K, where the proportion of fcc ice can be as high as 63.7 % due to these relatively low temperatures (Fig. S13). At lower temperatures (700 K and 800 K), coexistence of bcc and fcc appears (Fig. S14).

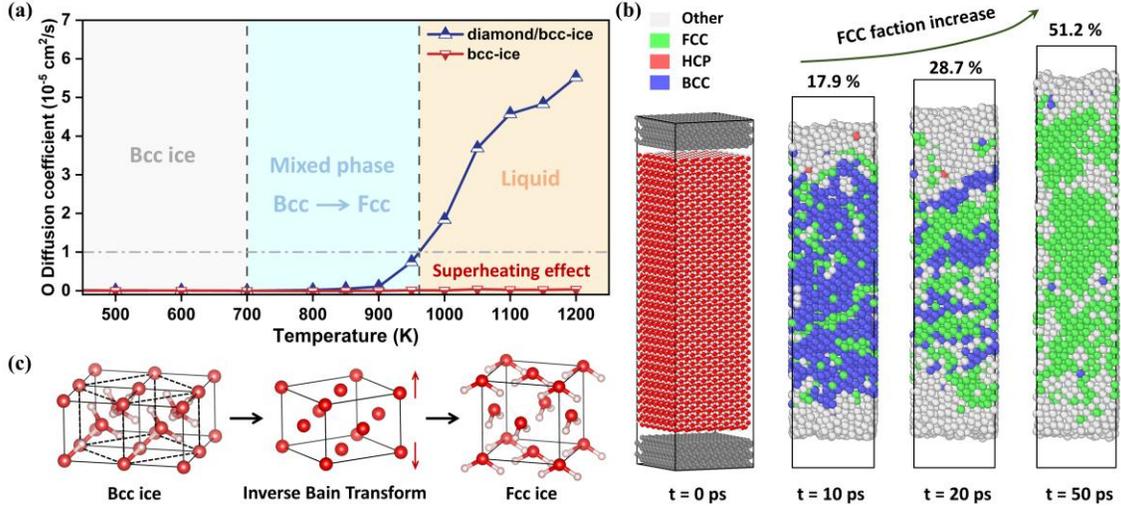

FIG. 3. Interface-dependent phase transitions. (a) Oxygen diffusion coefficients of the bcc-ice and diamond/bcc-ice systems and different phase regions. (b) The spontaneous phase transformation process from bcc ice to fcc ice in the diamond/bcc-ice system at 20 GPa and 900 K. Oxygen atoms are color-coded as green (fcc), blue (bcc), red (hcp) for ordered sublattices, and gray (disordered). (c) The inverse Bain transformation mechanism of cubic ice.

Further, the bcc to fcc transition mechanism can be understood from the structural correlation between the atomic planes in bcc and fcc structures. An fcc lattice precursor [middle panel of Fig. 4(c)] can simply derive from the oxygen bcc lattice, and an ideal fcc lattice can be acquired by expanding c-axis of the precursor, which is essentially an inverse Bain transformation. Driven by thermal effects, the oxygen atoms at high temperatures move upward from the equilibrium position in c-axis direction along with the bcc-fcc phase transition. Fcc ice can be considered as a transition phase during the melting process of bcc ice because the proportion of fcc ice decreases to 30% along with increasing disordered area and gradually melting at the interface for the situation of 950 K (Fig. S15), which explains why the stability fields of experimentally observed fcc ice phase are always near the melting line [12,13]. The coexistence of bcc and fcc phases is also observed in the other pressure ranges (from 25 GPa to 80 GPa), as shown in Fig. S16. It must be noted that the spontaneous phase transition from bcc ice to fcc ice under thermal effects has not been achieved in homogeneous solid ice, once again highlighting the necessity of introducing real interfaces to understand the phase behaviors of ice displayed in experiments. Additionally, we further extended the simulation system to 130 nm along the C-axis, constructing a model comprising 233,346 atoms and performed MD simulations at 20 GPa and 850 K. As presented in Figure S17, the ice was observed to transform gradually from bcc to fcc, initiating at the interface and propagating inwards. The transition was complete at approximately 1143 ps, which suggests that the mechanism can be effectively extended to more macroscopic systems.

## III. DISCUSSION

We compare a predicted phase diagram of ice with coupled interface effects with the results of relevant experimental and computational studies in Fig. 4. The phase diagram can be approximately delineated into four regions: solid bcc ice (light gray shading), superionic bcc ice (light blue shading), superionic fcc ice (light green shading), and liquid water (the white area). Additionally, we have consolidated relevant references on ice studies within our research scope— including melting data, bcc ice, and fcc ice. The melting curve of ice have been significant differences in experimental consensus. Reported melting temperatures at 20 GPa vary by up to 300 K, with this discrepancy increasing to 800 K at 40 GPa, as represented by the high and low melting data in Figure 4. Notably, the gap between these data is precisely spanned by the stability fields of the superionic bcc and fcc phases predicted in our work. Furthermore, our obtained melting lines (red solid line) are located within the higher experimental data measured by Queyroux et al. [27], Ahart et al. [24] and Schwager et al. [18,20] since 2004. It is also similar to recent computational results [37] using umbrella sampling simulations on superionic–liquid coexistence systems (brown dot dash line in Fig. 4), which proved that considering the real case - diamond/ice interface could approximate the experimental results and avoid complicated computational modeling. The lower experimental melting data intriguingly coincides with the superionic transition and the bcc-fcc phase boundaries identified in our study. This indicates the existence of superionic transition and bcc-fcc phase transition may cause the disappearance of some characteristic signals in Raman or x-ray diffraction measurements and further affect the determination of the melting in high pressure experiments. This may help explain why experiments have different observations of the melting curve of ice even at relatively low pressure, i.e., the lower melting line may correspond to the superionic transition of bcc ice or the transition from bcc to fcc ice rather than the real melting line. Schwager et al. observed an unidentified solid-solid phase transition line in the range of 20 GPa to 40 GPa (purple solid line in Fig. 4), which also further serves to confirm this hypothesis. [20].

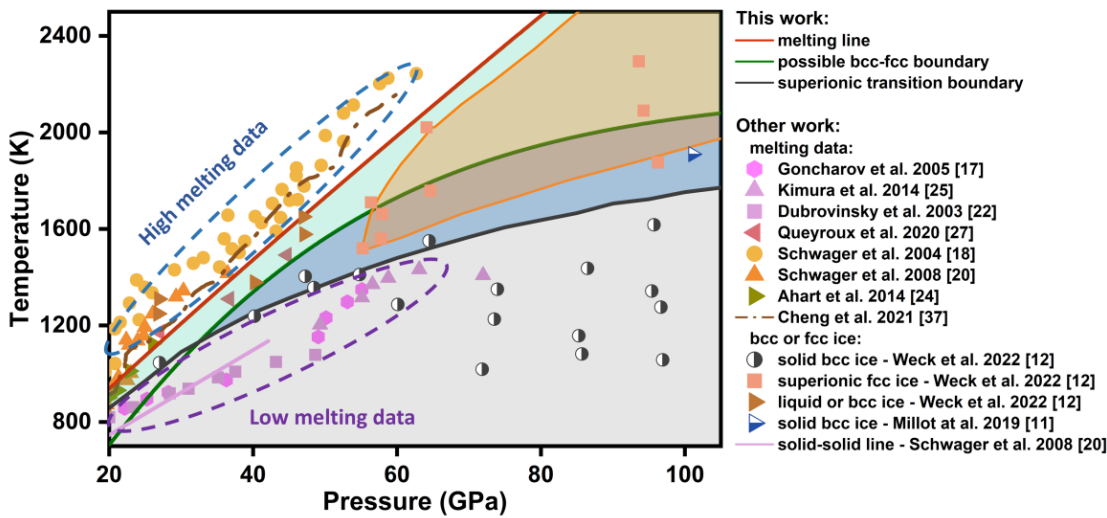

FIG 4. The high-pressure phase diagram of ice with a diamond/ice interface. The melting line (red solid line) was determined using Simon-Glatzel fits to oxygen diffusion coefficients. The superionic transition boundary of diamond/bcc-ice systems are shown in black solid line. The phase regions are color-coded as follows: light gray for solid bcc ice, light blue for superionic bcc ice, light green for possible superionic fcc ice, and white for liquid water. The experimental and theoretical data [11,12,17,18,20,22,24,25,27,37] from literature are also represented.

Regarding the bcc-to-fcc transition, at other pressures more than 20 GPa, we primarily observed a mixture of bcc and fcc ice phases. However, this coexistence also indicates the possible location of a bcc-fcc boundary (solid green line in Fig.4). It is noteworthy that the region between this boundary and the melting line shows consistency with the fcc stability intervals in the recent experiment reported by Weck et al. [12] in static compression experiments with XRD (orange shading in Fig. 4), proving the reliability of our results. The fcc ice phase we predict at low pressures (< 60 GPa) is vanishingly narrow (~100 K), which explains its absence in experimental reports given the inherent control challenges in temperature and pressure.

The previous calculations involving free energy have shown that fcc ice usually exists only at extremely high pressures, typically several hundred gigapascals [28,29]. Our molecular dynamics results and some experimental observations show that fcc ice indeed can be formed under relatively low pressure - dozens of gigapascals. The bcc-fcc phase transition we identify is a thermally driven process where oxygen atoms jump from the bcc equilibrium position to the fcc equilibrium position, accompanied by the expansion of the c-axis. During this process, the extra entropy of rapid diffusive hydrogen may improve the stability of this phase in thermodynamics. The predicted observation of fcc ice at 20 GPa also broadens the stability field of this phase beyond what has been reported experimentally to date [12,13,27].

Bcc plastic ice has recently been observed [38], and prior theoretical work [39] indicates that it can transform to fcc phase via heterogeneous nucleation in a mixed system consisting of glassy water and ice VII—a mechanism that resonates with our discussion on interfacial effects. Two-dimensional ice, which commonly exists on solid surfaces, in interlayers, or in confined pores, has a widely reported phase diagram [40,41]. Despite this, the influence of interfaces on its phase stability warrants deeper examination. Accounting for interface effects provides a new perspective for expanding the ice phase diagram. Beyond ice, other diamond/sample interfaces - such as diamond/Xe [42,43], diamond/hydrogen [44] or diamond/ammonia interfaces [45,46] - warrant further investigation into their potential superionic behaviors and phase transition mechanisms under extreme conditions. Introducing these interfaces to computational modeling can further clarify the novel or abnormal phenomena in relevant high-pressure experiments.

## IV. CONCLUSIONS

In summary, we extensively explored interfacial effects on superionic states, solid-solid phase transitions, and melting of bcc ice and proposed a phase diagram of ice with coupled interface effects. Our approach may be useful for examining a broad range of other materials studied in high-pressure experiments where the interface between the sample and components of the high-pressure device can induce transformations, affect physical and chemical properties, and potentially give rise to new phenomena associated with the low-dimensionality of the compressed interface.

The Supplemental Material contains Refs. [47-65].

## ACKNOWLEDGMENTS

Xing Meng is grateful to the National Natural Science Foundation of China (Grant No.12464035). Hanyu Liu acknowledges Scientific Research Innovation Capability Support Project for Young Faculty (Grant No. ZYGXONJSKYCXNLZCXM-M12), National Natural Science Foundation of China (Grant No. 52288102, 52090024, and T2494523011), Program for Jilin University Science and Technology Innovative Research Team (2021TD-05), the Fundamental Research Funds for the Central Universities and computing facilities at the High-Performance Computing Centre of Jilin University. Pengfei Hou thanks the National Natural Science Foundation of China (Grant No.124B2073). Russell J. Hemley acknowledges support from the U.S. National Science Foundation (DMR-2104881). The computational resource is provided by the Open Source Supercomputing Center of S-A-I.

## DATA AVAILABILITY

The data supporting the findings of this study are available within the paper, and a detailed description of the calculations is included in the Supplemental Material. All original data generated for the study and the machine learning potential for high-pressure diamond/ice interface constructed in this study are in the Supplemental Material and available from the corresponding author upon reasonable request.

**Reference**

[1] L. I. Cleeves, E. A. Bergin, C. M. O. D. Alexander, F. Du, D. Graninger, K. I. Öberg, and T. J. Harries, Science **345**, 1590 (2014).
[2] O. Tschauner *et al.*, Science **359**, 1136 (2018).
[3] R. Z. Ma *et al.*, Nature **577**, 60 (2020).
[4] R. J. Hemley, L. C. Chen, and H. K. Mao, Nature **338**, 638 (1989).
[5] R. J. Hemley, A. P. Jephcoat, H. K. Mao, C. S. Zha, L. W. Finger, and D. E. Cox, Nature **330**, 737 (1987).
[6] M. Somayazulu, J. Shu, C.-s. Zha, A. F. Goncharov, O. Tschauner, H. K. Mao, and R. J. Hemley, The Journal of Chemical Physics **128**, 064510 (2008).
[7] T. Bartels-Rausch *et al.*, Reviews of Modern Physics **84**, 885 (2012).
[8] C. G. Salzmann, The Journal of Chemical Physics **150**, 060901 (2019).
[9] T. C. Hansen, Nature Communications **12**, 3161 (2021).
[10] P. Loubeyre, R. LeToullec, E. Wolanin, M. Hanfland, and D. Hausermann, Nature **397**, 503 (1999).
[11] M. Millot, F. Coppari, J. R. Rygg, A. Correa Barrios, S. Hamel, D. C. Swift, and J. H. Eggert, Nature **569**, 251 (2019).
[12] G. Weck, J.-A. Queyroux, S. Ninet, F. Datchi, M. Mezouar, and P. Loubeyre, Physical Review Letters **128**, 165701 (2022).
[13] V. B. Prakapenka, N. Holtgrewe, S. S. Lobanov, and A. F. Goncharov, Nature Physics **17**, 1233 (2021).
[14] J.-A. Hernandez and R. Caracas, Physical Review Letters **117**, 135503 (2016).
[15] R. Caracas, Physical Review Letters **101**, 085502 (2008).
[16] J.-A. Hernandez and R. Caracas, The Journal of Chemical Physics **148**, 214501 (2018).
[17] A. F. Goncharov, N. Goldman, L. E. Fried, J. C. Crowhurst, I. F. W. Kuo, C. J. Mundy, and J. M. Zaug, Physical Review Letters **94**, 125508 (2005).
[18] B. Schwager, L. Chudinovskikh, A. Gavriliuk, and R. Boehler, Journal of Physics: Condensed Matter **16**, S1177 (2004).
[19] J.-F. Lin, E. Gregoryanz, V. V. Struzhkin, M. Somayazulu, H.-k. Mao, and R. J. Hemley, Geophysical Research Letters **32**, L11306, L11306 (2005).
[20] B. Schwager and R. and Boehler, High Pressure Research **28**, 431 (2008).
[21] F. Datchi, P. Loubeyre, and R. LeToullec, Physical Review B **61**, 6535 (2000).
[22] N. Dubrovinskaia and L. and Dubrovinsky, High Pressure Research **23**, 307 (2003).
[23] M. R. Frank, Y. Fei, and J. Hu, Geochimica et Cosmochimica Acta **68**, 2781 (2004).
[24] M. Ahart, K. Amol, G. Stephen, B. Reinhard, and R. J. and Hemley, High Pressure Research **34**, 327 (2014).
[25] T. Kimura, Y. Kuwayama, and T. Yagi, The Journal of Chemical Physics **140**, 074501 (2014).
[26] J.-F. Lin, B. Militzer, V. V. Struzhkin, E. Gregoryanz, R. J. Hemley, and H.-k. Mao, The Journal

of Chemical Physics **121**, 8423 (2004).
[27] J. A. Queyroux *et al.*, Physical Review Letters **125**, 195501 (2020).
[28] H. F. Wilson, M. L. Wong, and B. Militzer, Physical Review Letters **110**, 151102 (2013).
[29] J. Sun, B. K. Clark, S. Torquato, and R. Car, Nature Communications **6**, 8156 (2015).
[30] W. L. Vos, L. W. Finger, R. J. Hemley, and H.-k. Mao, Physical Review Letters **71**, 3150 (1993).
[31] Y.-H. Wang *et al.*, Nature **600**, 81 (2021).
[32] Y. Tian *et al.*, Science **377**, 315 (2022).
[33] M.-S. Miao and R. Hoffmann, Accounts of Chemical Research **47**, 1311 (2014).
[34] See Supplemental Material at http://link.aps.org/ supplemental/XXX for further details and discussions.
[35] C. Niu, H. Zhang, J. Zhang, Z. Zeng, and X. Wang, The Journal of Physical Chemistry Letters **13**, 7448 (2022).
[36] A. B. Belonoshko, N. V. Skorodumova, A. Rosengren, R. Ahuja, B. Johansson, L. Burakovsky, and D. L. Preston, Physical Review Letters **94**, 195701 (2005).
[37] B. Cheng, M. Bethkenhagen, C. J. Pickard, and S. Hamel, Nature Physics **17**, 1228 (2021).
[38] M. Rescigno *et al.*, Nature **640**, 662 (2025).
[39] M. J. Zimoń and F. Martelli, The Journal of Chemical Physics **158**, 114501 (2023).
[40] V. Kapil, C. Schran, A. Zen, J. Chen, C. J. Pickard, and A. Michaelides, Nature **609**, 512 (2022).
[41] J. Jiang, Y. Gao, L. Li, Y. Liu, W. Zhu, C. Zhu, J. S. Francisco, and X. C. Zeng, Nature Physics **20**, 456 (2024).
[42] E. Kim, M. Nicol, H. Cynn, and C. S. Yoo, Physical Review Letters **96**, 035504, 035504 (2006).
[43] A. P. Jephcoat, H. k. Mao, L. W. Finger, D. E. Cox, R. J. Hemley, and C. s. Zha, Physical Review Letters **59**, 2670 (1987).
[44] C. S. Zha, Z. X. Liu, and R. J. Hemley, Physical Review Letters **108**, 146402, 146402 (2012).
[45] C. J. Pickard and R. J. Needs, Nature Materials **7**, 775 (2008).
[46] A. Mondal, R. J. Husband, H. P. Liermann, and C. Sanchez-Valle, Physical Review B **107**, 224108 (2023).
[47] M. Bernetti and G. Bussi, The Journal of Chemical Physics **153**, 114107 (2020).
[48] M. Ceriotti, M. Parrinello, T. E. Markland, and D. E. Manolopoulos, The Journal of Chemical Physics **133**, 124104 (2010).
[49] I. R. Craig and D. E. Manolopoulos, Journal of Chemical Physics **121**, 3368 (2004).
[50] Z. Fan, W. Chen, V. Vierimaa, and A. Harju, Computer Physics Communications **218**, 10 (2017).
[51] S. Grimme, J. Antony, S. Ehrlich, and H. Krieg, The Journal of Chemical Physics **132**, 154104 (2010).
[52] S. Grimme, S. Ehrlich, and L. Goerigk, Journal of Computational Chemistry **32**, 1456 (2011).
[53] G. Kresse and J. Furthmüller, Computational Materials Science **6**, 15 (1996).
[54] G. Kresse and D. Joubert, Physical Review B **59**, 1758 (1999).
[55] T. D. Kühne *et al.*, Journal of Chemical Physics **152**, 194103, 194103 (2020).
[56] N. Michaud-Agrawal, E. J. Denning, T. B. Woolf, and O. Beckstein, Journal of Computational Chemistry **32**, 2319 (2011).
[57] K. Momma and F. Izumi, Journal of Applied Crystallography **44**, 1272 (2011).
[58] S. Nosé, The Journal of Chemical Physics **81**, 511 (1984).
[59] J. P. Perdew, K. Burke, and M. Ernzerhof, Physical Review Letters **77**, 3865 (1996).
[60] S. Plimpton, Journal of Computational Physics **117**, 1 (1995).
[61] K. Song *et al.*, Nature Communications **15**, 10208 (2024).
[62] A. Stukowski, Modelling and Simulation in Materials Science and Engineering **18**, 015012, 015012 (2010).
[63] H. Wang, L. Zhang, J. Han, and W. E, Computer Physics Communications **228**, 178 (2018).
[64] V. Wang, N. Xu, J.-C. Liu, G. Tang, and W.-T. Geng, Computer Physics Communications **267**, 108033 (2021).
[65] Y. Zhang, H. Wang, W. Chen, J. Zeng, L. Zhang, H. Wang, and W. E, Computer Physics Communications **253**, 107206 (2020).

**Supplementary Material for**

Interface-dependent Phase Transitions and Ultrafast Hydrogen Superionic Diffusion of $H_2O$ Ice

Pengfei Hou[1], Yumiao Tian[1], Zifeng Liu[1], Junwen Duan[1], Hanyu Liu[1, 4, *], Xing Meng[1, 2, †], Russell J. Hemley[3], Yanming Ma[1, 4, 5, ‡]

[1]*Key Laboratory of Material Simulation Methods and Software of Ministry of Education, State Key Laboratory of High Pressure and Superhard Materials, College of Physics, Jilin University, Changchun 130012, China.*

[2]*Key Laboratory of Physics and Technology for Advanced Batteries (Ministry of Education), College of Physics, Jilin University, Changchun 130012, China.*

[3]*Department of Physics, Department of Chemistry, and Department of Earth and Environmental Sciences, University of Illinois Chicago, Chicago, IL 60607.*

[4]*International Center of Future Science, Jilin University, Changchun 130012, China.*

[5]*School of Physics, Zhejiang University, Hangzhou, China.*

[*]Contact author: hanyuliu@jlu.edu.cn
[†]Contact author: mengxing@jlu.edu.cn
[‡]Contact author: mym@jlu.edu.cn

**This file includes:**

   Supplementary Material text
   Supplemental Figures: Figs. S1 to S17
   Supplemental References

## Supplementary Material Text
**Interface reconstruction between bcc ice and diamond.**

In diamond-anvil cell experiments on ice, water samples are typically placed between two diamond anvils that naturally form the diamond/ice interface when compressed. Choosing the common bcc-based phase ice VII, we first constructed an interface model - diamond (100) /bcc ice (100) containing eight carbon layers as the diamond region and ten ice layers within the bcc ice, followed by geometry optimization at 100 GPa. The results of the optimization (Fig. S1) revealed that water molecules close to the surface dissociated and hydrogen or oxygen bonded with carbon under high pressure but maintained the typical bcc lattice structure. The interface model thus maintains the stability of bcc oxygen sublattice in the 10 ps AIMD simulations of 1000 K, 1500 K, and 2000 K, respectively, in the canonical (NVT) ensemble (Fig. S1).

**The training process of NNP by active learning.**

To overcome size effects in standard first-principles simulations, we constructed a neural network potential (NNP) for diamond, bcc ice, and their prevalent interface models diamond (100) /bcc ice (100) over a wide P-T range using active learning scheme. The iterative concurrent active learning scheme executed by the Deep Potential Generator (DP-GEN) [1] and DeePMD-kit [2] is utilized to check the accuracy of the NNP in real time and effectively guarantee the quality of the acquired NNP by adding DFT results of the candidate snapshots into dataset. The entire flowchart is shown in the left panel of Fig. S2. The whole active learning process starts with a small dataset of 1,511 snapshots that are sampled at intervals of 100 fs in AIMD simulations on initial models (atomic structures in Fig. S2) comprising diamond, bcc ice, and their interface configurations with varying numbers of ice layers (from 2 layers to 12 layers, noted as C144-32$H_2O$, C144-64$H_2O$, C144-96$H_2O$, C144-128$H_2O$, C144-160$H_2O$, and C144-192$H_2O$ respectively), ensuring the NNP can effectively capture the thickness-dependent evolution of physical properties of ice. The whole active learning process could be divided into three stages: training, exploration, and correction. In the training stage, two NNPs are constructed using the DeePMD-kit scheme based on the random initialization parameters. And then NPT MD explorations are carried out on the above initial models based on the trained NNPs under different thermodynamic conditions. The explorations are classified into two groups: the high pressure (group I) and the low pressure (group II). Group I includes states in the range 50 GPa < P < 120 GPa with 500 K < T < 2000 K. Group II includes states in the range 20 GPa < P < 50 GPa with 500 K < T < 1200 K. The total exploration time is ranging from 500 fs in the first iteration to 100000 fs in the last iteration in a 1 fs step width. For correction stage, we evaluated the deviations of different NNPs according to the maximal atomic force deviation ($\sigma_f^{max}$) and label the MD snapshots as accurate, candidate or failed based on $\sigma_f^{max} < \sigma_l$, $\sigma_l \leq \sigma_f^{max} < \sigma_h$, or $\sigma_f^{max} \geq \sigma_h$, where $\sigma_l$ and $\sigma_h$ correspond to the lowest and highest trust levels of atomic force deviation, respectively. And we calibrated the NNPs by adding the DFT results from the candidate snapshots into dataset. The iterative active learning process concludes once almost all snapshots are labeled as accurate. Through dynamic adjustments of the force threshold ($\sigma_l$ and $\sigma_h$), ultimately the $\sigma_f^{max}$ of almost all evaluated structures (> 98 %) in the last active learning iteration are less than 10% and their average value is only 5.03 % [Fig. S3 (a)]. In fact, the average atomic force deviations [Fig. S3 (b)] are only 1.06 %, indicating that the current dataset is sufficient to contain the information of the systems under the given temperature and pressure conditions. Eventually, a high-quality dataset containing 17076 pieces of DFT data was collected from about 66.41 million snapshots after 13 iterations. The distribution of the training dataset in the 2D principal component space of the descriptor is shown in Fig. S2.

**The NNP accuracy test.**

Considering the high efficiency of the recently Graphics Processing Units Molecular Dynamics (GPUMD) package [3], we trained a NNP based on neuroevolution potential (NEP) [4] approach using the above training dataset for conducting subsequent large-scale MD simulations. During the entire training process of 300,000 steps, as shown in Fig. S4, the root-mean-square errors (RMSE) of energy, force and virial gradually decreased and eventually converged, accompanied by an increase in the regularization parameter $L_2$ first and then a decrease, indicating that no overfitting occurred in the training process. And then we randomly collected total 1120 atomic images from all snapshots in the last active learning iteration and performed the DFT calculations on these atomic images. The errors of the NNP based on the NEP approach in predicting the atomic energy, virial, and force are compared with the DFT results in Fig. S5. For training dataset and testing dataset, the predictions of NNP both show a good diagonal relationship (y = x) with the actual value from DFT calculations. Particularly, we have noticed that the force RMSE (346.2 meV/Å) in the training set are larger than that of (222.9 meV/Å) testing dataset because of sampled "non-physical" structures with larger atomic forces in the initial active sampling process due to the poor stability of initial NNP. In addition, we also randomly added some structures under more extreme conditions (up to 140 GPa and 5000 K) in the initial stage. These non-physical/extreme structures can enhance the stability of the NNP to some extent and thus are retained in the training dataset.

**The parameter setting of DFT calculation.**

We perform all DFT calculations of single-point energy to obtain the energy, force and Virial information of the sampled structures by using the projector augmented wave (PAW) method [5] applied in the Vienna Ab Initio Simulation Package (VASP) [6]. The exchange correlation effect is expressed through the generalized gradient approximation (GGA) of the Perdew–Burke–Emzerhof (PBE) method [7]. The Grimme's D3 methods are used to consider vdW corrections for all DFT calculations [8,9]. The cut-off energy and the energy convergence criterion are 900 eV and $10^{-6}$ eV, respectively. The smallest allowed spacing between k points is 0.5 Å$^{-1}$ for all initial models. The flag PREC is set to Accurate for more accurate energy and force. The NVT AIMD simulation for constructing initial rough training dataset were performed on initial models using the CP2K package [10] with PBE methods, Grimme's D3 corrections and orbital transformation method. The flag CUTOFF and REL_CUTOFF are set to 500 and 50 Ry, respectively.

**DP-GEN and DeePMD-kit setup.**

The embedding nets and the fitting nets for the NNPs are (25, 50, 100) and (240, 240, 240) orderly. The cut-off radius is 6 Å with the 0.5 Å smoothing parameter (rcut_smth). The prefactors of the energy, the force, and the virial terms in the loss functions vary from 0.02 to 1, from 1000 to 1, and from 0.01 to 0.1. The starting learning rate is 0.001 and exponentially decayed at the training process. Lammps [11] is used to perform the MD exploration process in the iterations. For the MD exploration, the total exploration time is ranging from 500 fs in the first iteration to 100000 fs in the last iteration. NPT MD simulations are performed on the structures at different thermodynamic conditions - the high pressure (group I) and the low pressure (group II). Group I includes states in the range 50 GPa < P < 120 GPa with 500 K < T < 2000 K. Group II includes states in the range 20 GPa < P < 50 GPa with 500 K < T < 1200 K. The above NPT ensembles are implemented by the Nose-Hoover thermostat and barostat [12]. The time step of all simulations is 1fs. During the MD simulation, all atoms moved freely in all directions.

**GPUMD-NEP setup.**

We used GPUMD v3.9.5 [3] to train a NNP based on fourth-generation NEP approach [4]. The cutoff radii for radial and angular descriptor parts are 5 Å and 3 Å, respectively. For both the radial and the angular descriptor components, we used 4 radial functions constructed from a linear combination of basis functions. The descriptor vector for one element thus has 30 components. There are 50 neurons used in the hidden layer and the NN architecture for each element can be written as 30-50-1, corresponding to total 5611 optimized parameters. The training was performed with a batch size of 1000 structures for 300,000 generations(steps) on 4090 GPUs.

To investigate the large-scale dynamic behaviors of bcc-ice (6,000 atoms) and diamond/bcc-ice (20,648 atoms) systems, we performed 50 ps NPT MD simulation on the two model under different thermodynamic conditions (from 20 GPa to 120 GPa, from 400 K to 2100 K). To explore the influence of simulation time on the interface effect, we also performed 500 ps and 1000 ps NPT MD simulations on the two models at 100 GPa with 1700 K, 1800 K, 1900 K, 2000 K, and 2100 K. In bcc-fcc phase transition part, we performed 100 ps NPT MD simulation at 20 GPa and 900 K to prove the alternating evolution of the disordered region and the fcc region (Fig. S11) and 600 ps cooling NPT MD simulation with 1 K/ps from 900 K to 300 K at 20 GPa to confirm the reversibility of phase transitions (Fig. S12). Exceptionally, a complete bcc-fcc phase transition also occurred under the conditions of 20 GPa and 850 K during the 350 ps NPT MD simulation (Fig. S13). All NPT MD simulations apply the the stochastic cell rescaling method [13] and their time steps are 0.1 fs for the fineness of the simulation. Relevant properties including the mean square displacement (MSD) and diffusion coefficient (D) of each element are calculated from 20 ps NVE MD simulation after NPT simulations. In particular, in order to eliminate the influence of nuclear quantum effects on our results, we performed 50 ps path-integral molecular dynamics simulations (PIMD) [14] and then calculated MSD and D from 20 ps ring-polymer molecular dynamics simulations (RPMD) [15] at 100 GPa with 900 K, 1200 K, 1400 K, 1600 K, 1800 K, 2000 K, 2100 K. The number of beads is set to 16 and time step is also 0.1 fs.

**Other data analysis.**

The MDAnalysis [16] package is used to read the atomic trajectories from MD simulation for data analysis including hydrogen rotation rate and the projections of oxygen and hydrogen positions along O-H···O triplets [Fig. 2(c)]. The related modeling and crystal visualization also borrowed the VASPKIT [17] and VESTA package [18]. Bcc, fcc and other sublattices of ice were identified using interval common neighbor analysis with variable cutoff, as implemented in the OVITO package [19]. Oxygen atoms are color-coded as green (fcc), blue (bcc), red (hcp) for ordered sublattices, and gray (disordered).

**Supplemental Figures.**

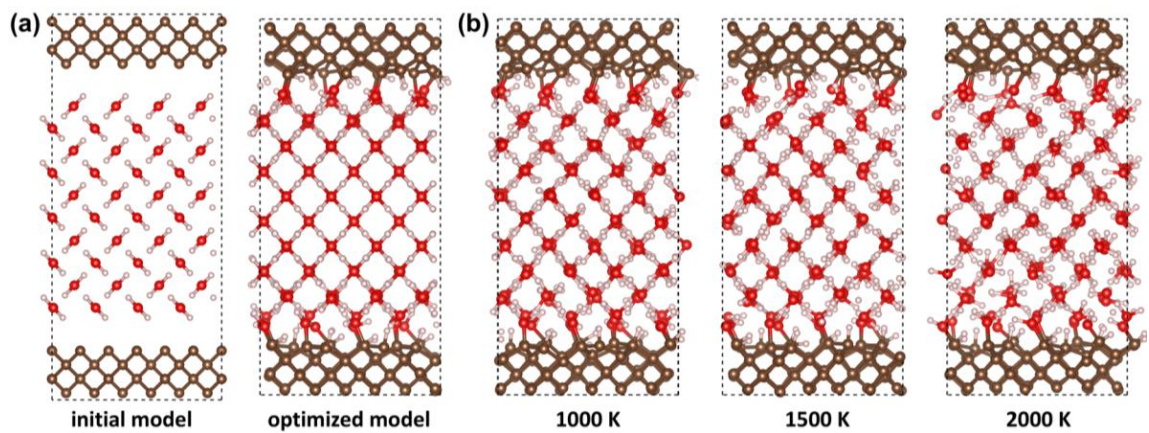

**FIG. S1.** Interface reconstruction between bcc ice and diamond. (a) Initial interface model of diamond (100) /bcc ice (100) containing eight carbon layers and ten bcc ice layers and optimized model at 100 GPa. (b) Final snapshots of the interface model after 10 ps NVT AIMD simulations of 1000 K, 1500 K and 2000 K.

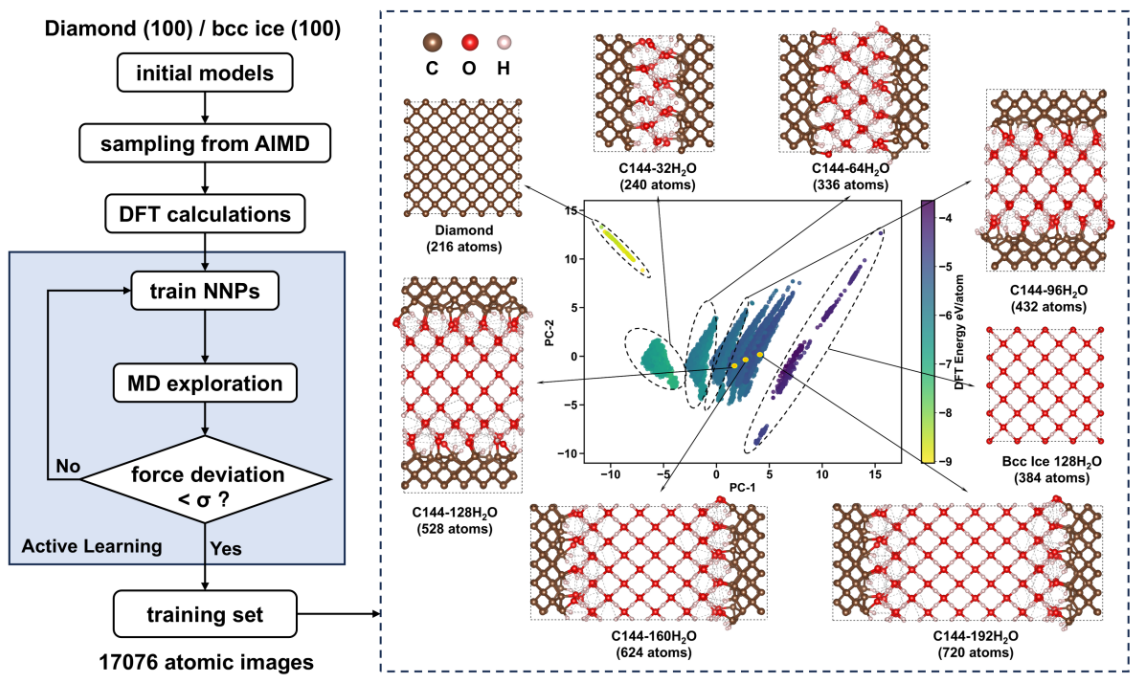

**FIG. S2.** The flowchart in developing the NNP for diamond (100) /bcc ice (100) interface system with the active learning scheme and the distribution of the training dataset in the 2D principal component space of the descriptor. The initial structures for conducting the MD exploration and the corresponding chemical compositions are also listed.

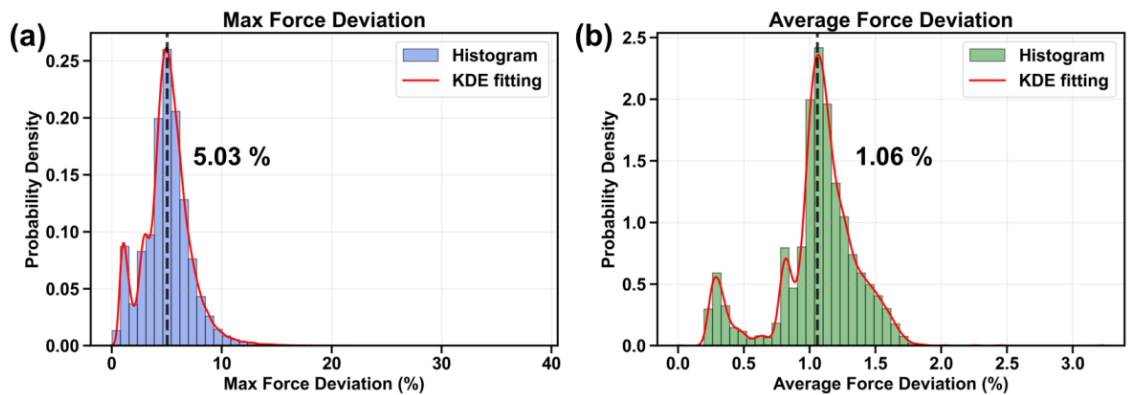

**FIG. S3.** The maximum atomic force deviation (a) and average atomic force deviation (b) of the explored structures in the last active learning iteration.

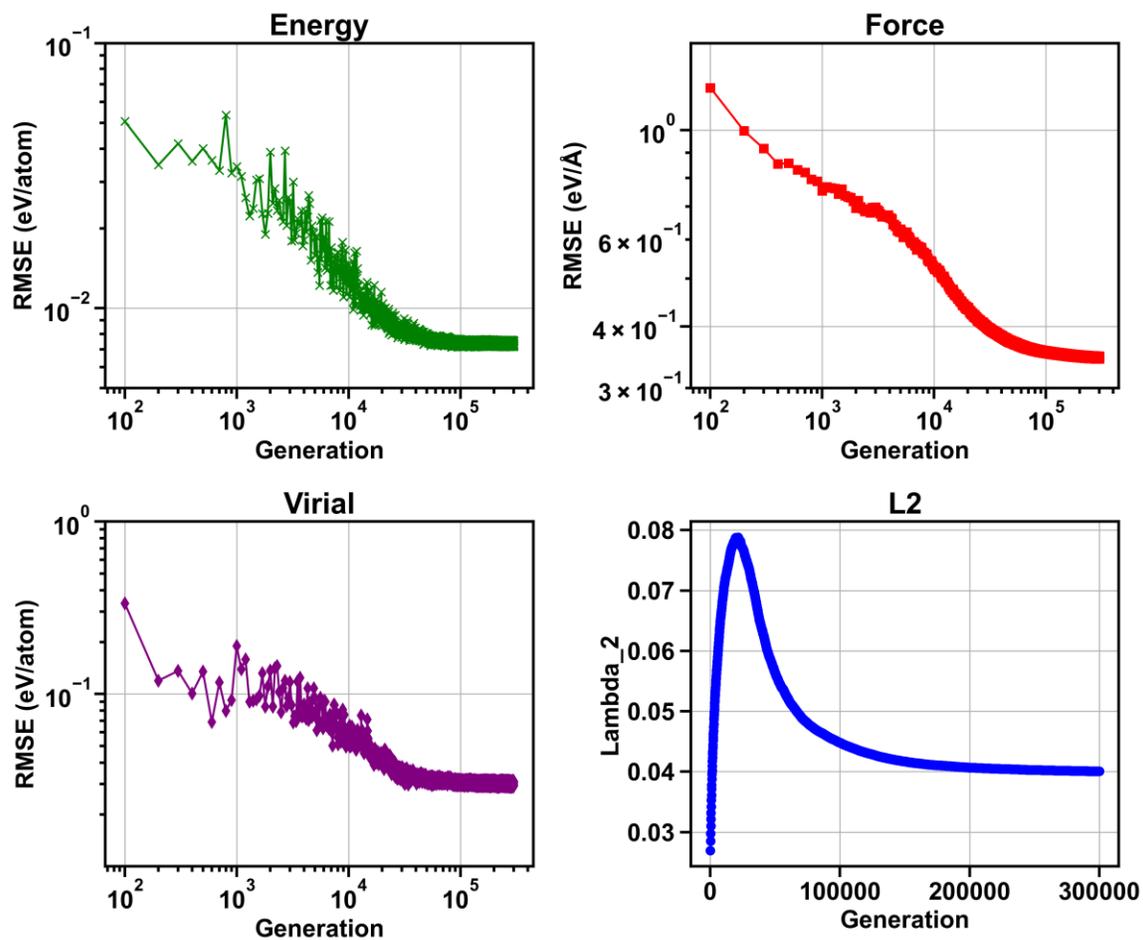

**FIG. S4.** The convergence of NNP on (a) energy, (b) atomic force, (c) virial and (d) regularization parameter $L_2$ after 300,000 training steps using the neural evolution algorithms on the training dataset.

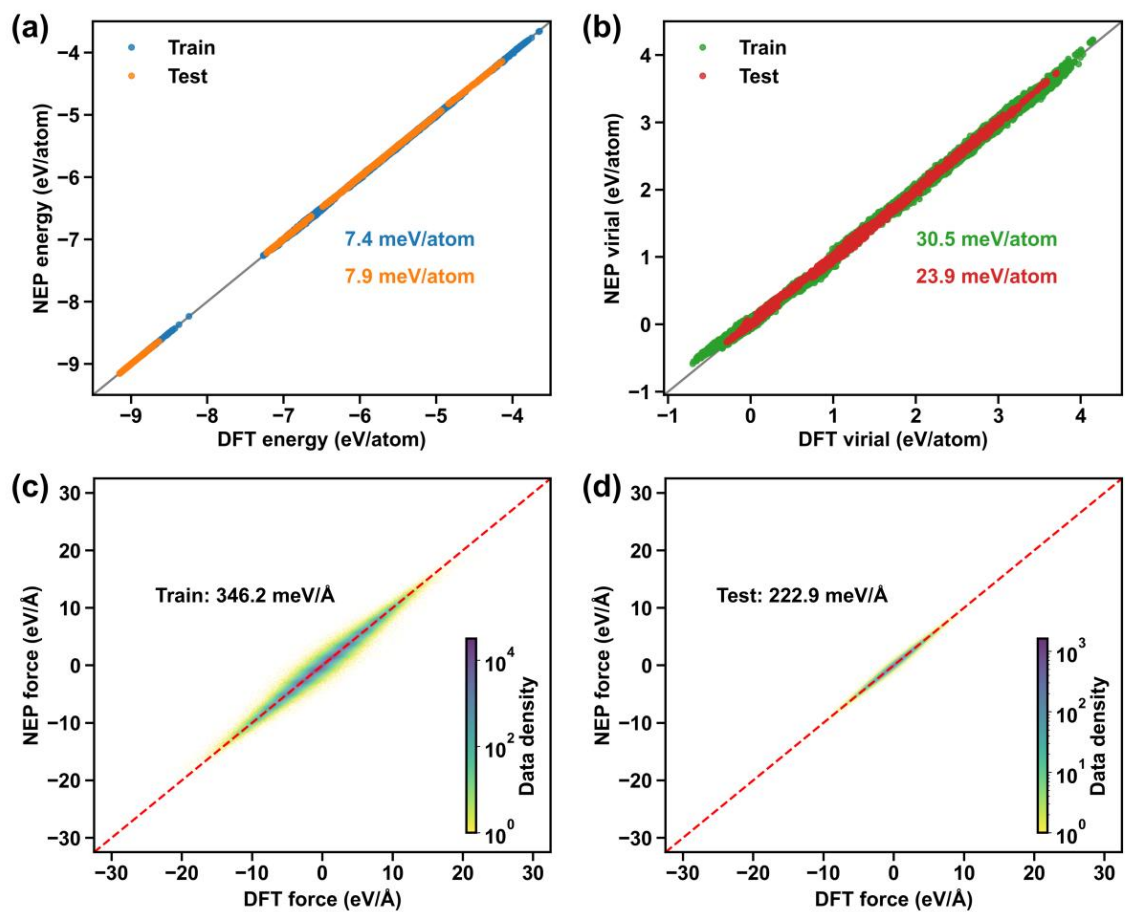

**FIG. S5.** The diagnostic plots between DFT and NNPs for (a) energy and (b) virial on training dataset and testing dataset. The diagnostic plots and distribution between DFT and NNPs for the force on (c) training dataset and (d) testing dataset.

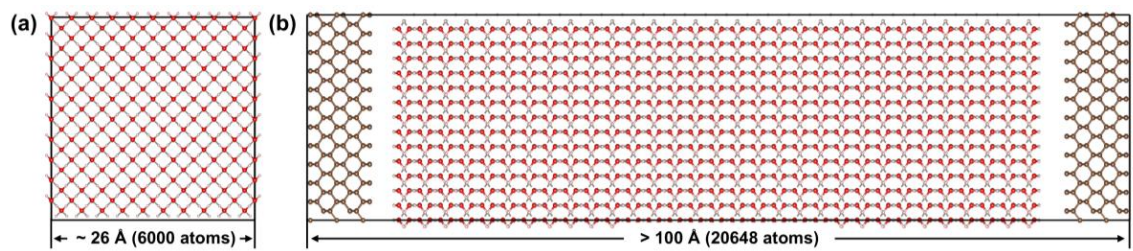

**FIG. S6.** Large-scale models for subsequent MD simulations: (a) bcc ice model (2000 H$_2$O molecules, 6000 atoms total) and (b) a diamond (100)/bcc ice (100) interface model (2420 C atoms, 6076 H$_2$O molecules, 20,648 atoms total).

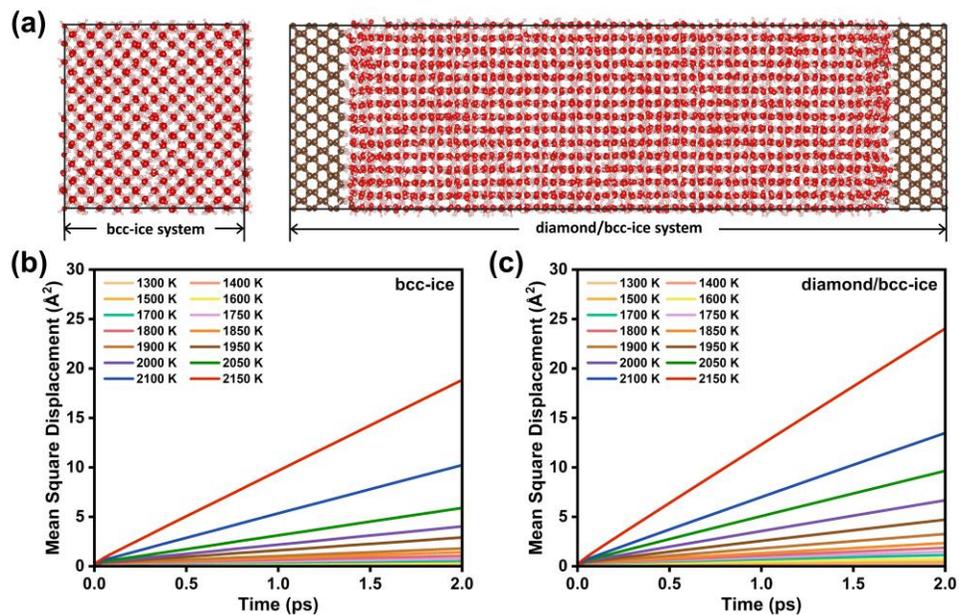

**FIG. S7.** (a) Final snapshots of bcc ice and diamond (100) /bcc ice (100) interface after 50 ps NPT MD simulation at 100 GPa and 2000 K. Mean square displacements of the (b) bcc ice and (c) diamond (100) /bcc-ice (100) systems at different temperatures.

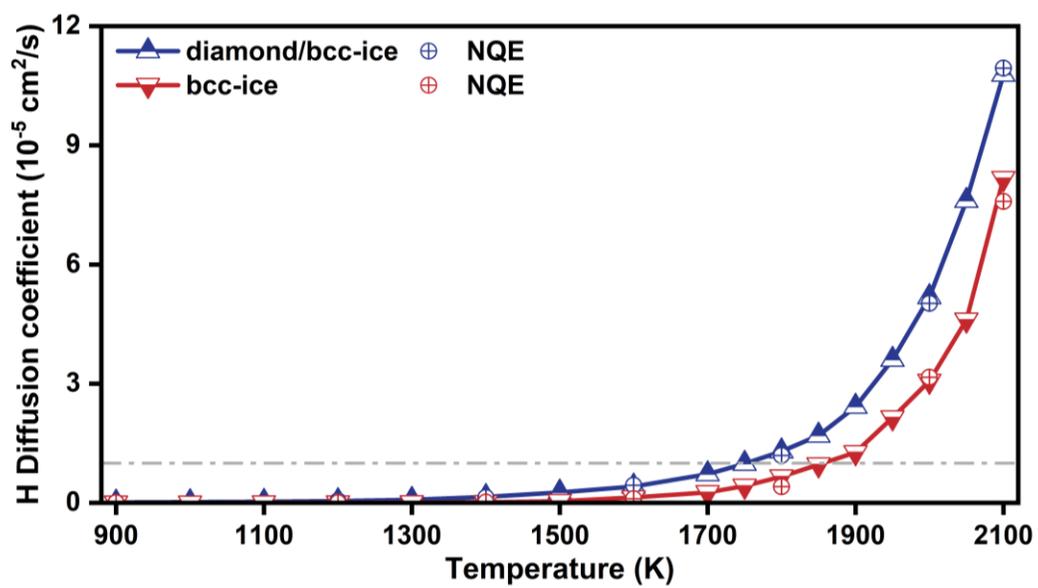

**FIG. S8.** Hydrogen diffusion coefficients for the two models with and without nuclear quantum effects from 900 K to 2100 K.

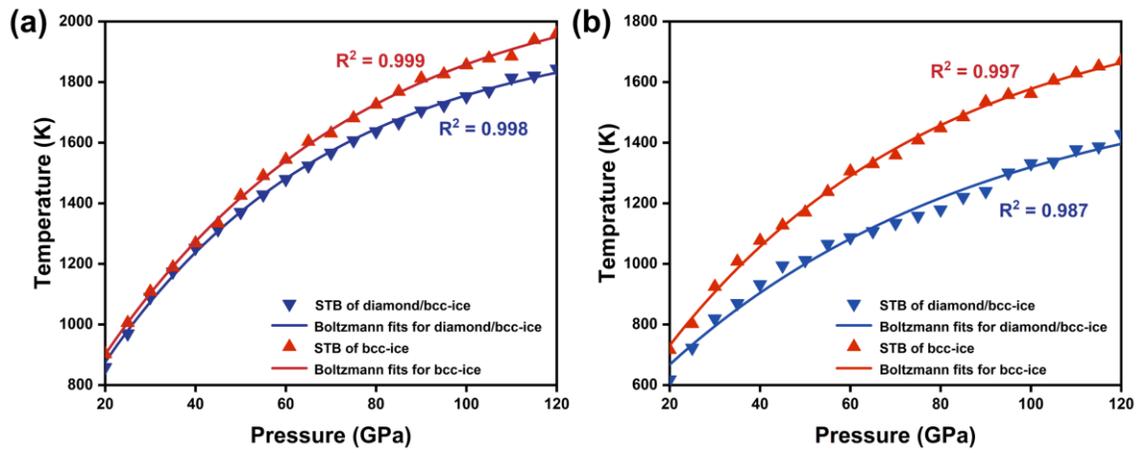

**FIG. S9.** The calculated superionic transition boundary (STB) of bcc-ice and diamond/bcc-ice systems with varying pressures from 20 GPa to 120 GPa with (a) $10^{-5}$ cm$^2$/s and (b) $10^{-6}$ cm$^2$/s as thresholds, respectively. The STB for bcc-ice and diamond/bcc-ice systems was obtained through Boltzmann fits to hydrogen diffusion coefficients.

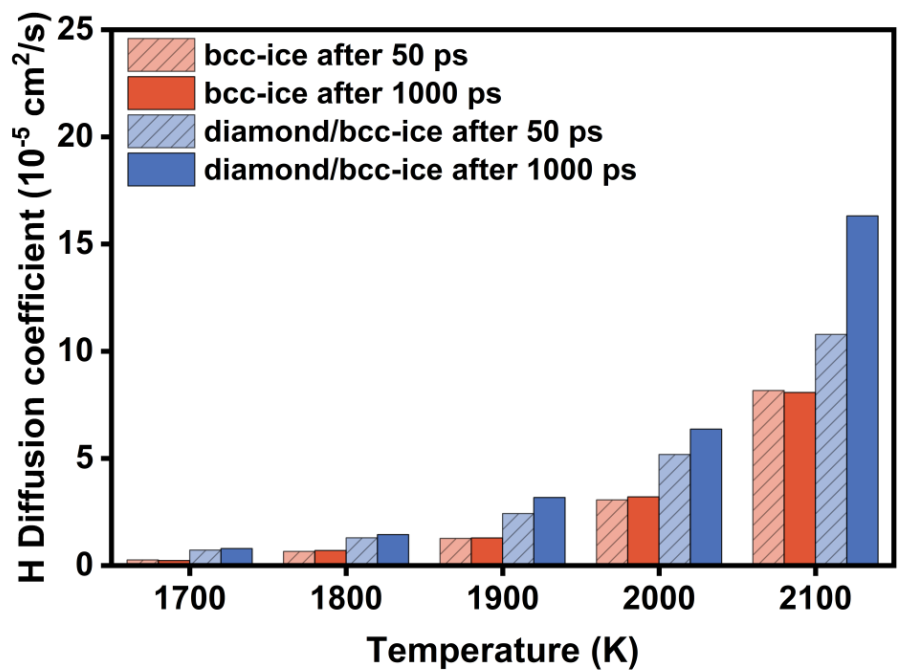

**FIG. S10.** The influence of simulation time on the hydrogen diffusion of the bcc-ice and diamond/bcc-ice systems.

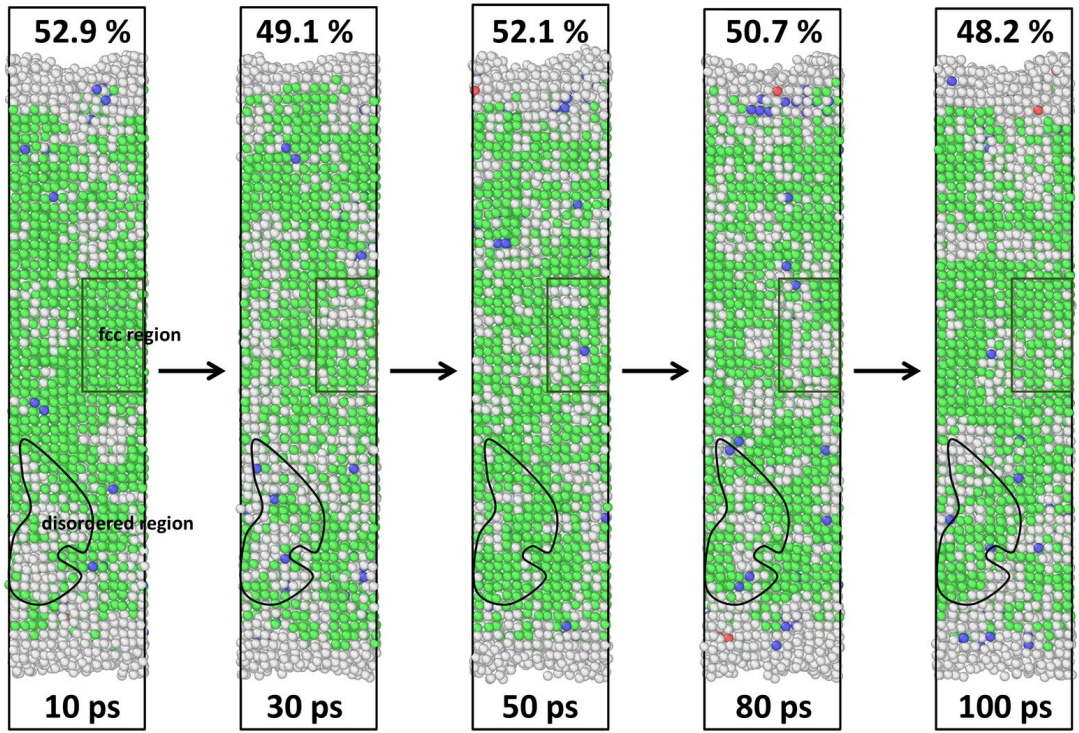

**FIG. S11.** Snapshots of 100 ps equilibrium MD simulation (20 GPa, 900 K) for diamond/bcc-ice system that have undergone bcc-fcc phase transition and the corresponding distribution and proportion of fcc ice. Oxygen atoms are color-coded as: green (fcc), blue (bcc), red (hcp) for ordered sublattices, and gray (disordered).

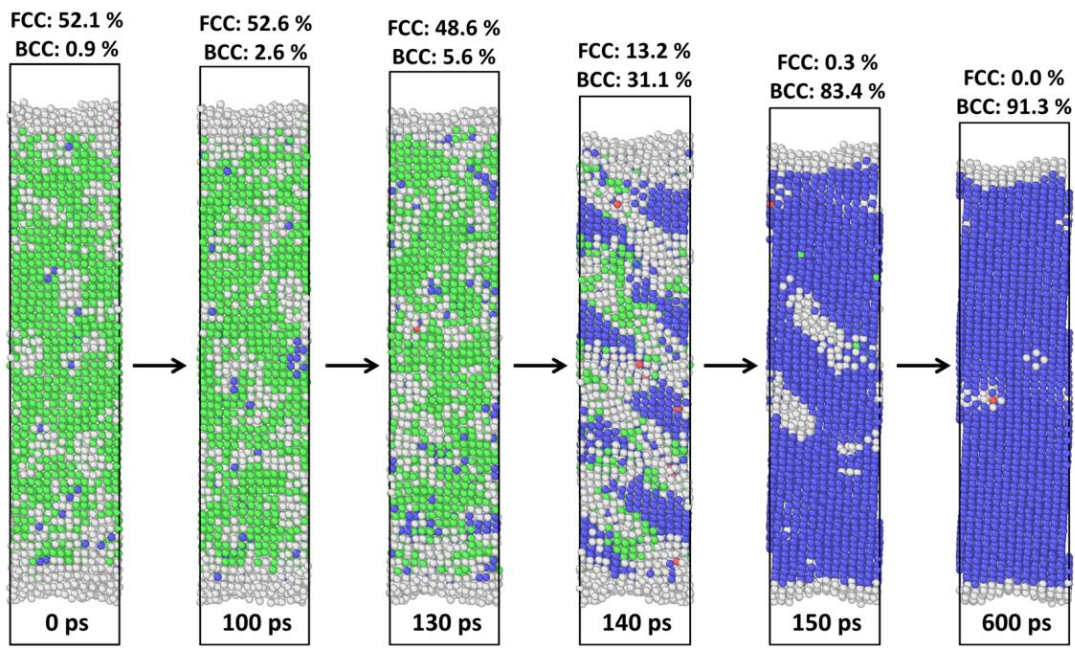

**FIG. S12.** Snapshots of MD simulation for diamond/bcc-ice system cooled from 900 K to 300 K at a rate of 1 K/ps (20 GPa) and the corresponding distribution and proportion of fcc and bcc ice. Oxygen atoms are color-coded as: green (fcc), blue (bcc), red (hcp) for ordered sublattices, and gray (disordered).

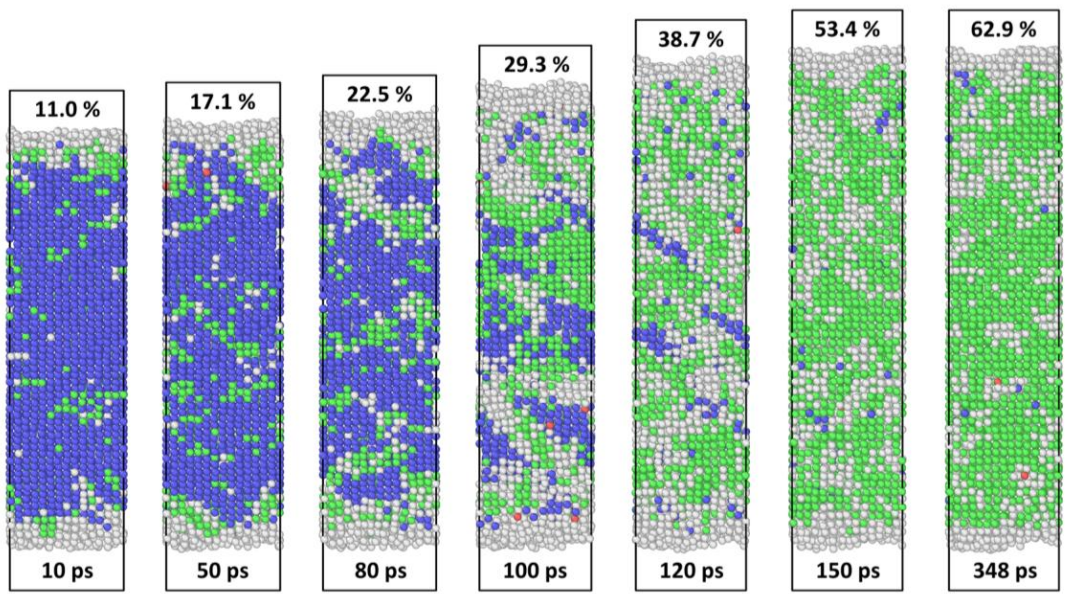

**FIG. S13.** The spontaneous phase transformation process from bcc ice to fcc ice in the diamond/bcc-ice system at 20 GPa and 850 K. Oxygen atoms are color-coded as: green (fcc), blue (bcc), red (hcp) for ordered sublattices, and gray (disordered).

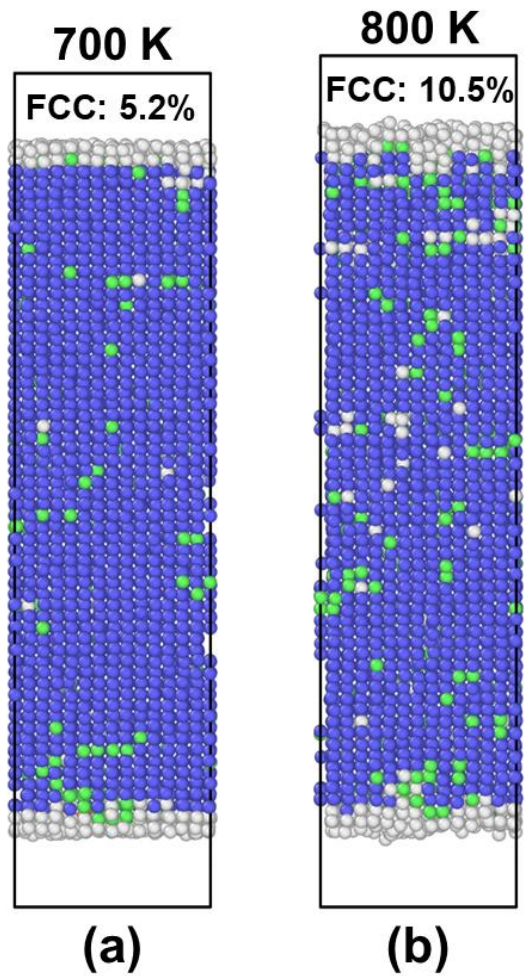

**FIG. S14.** The coexist phases of bcc and fcc at the temperature of (a) 700 K and (b) 800 K. Oxygen atoms are color-coded as: green (fcc), blue (bcc), red (hcp) for ordered sublattices, and gray (disordered).

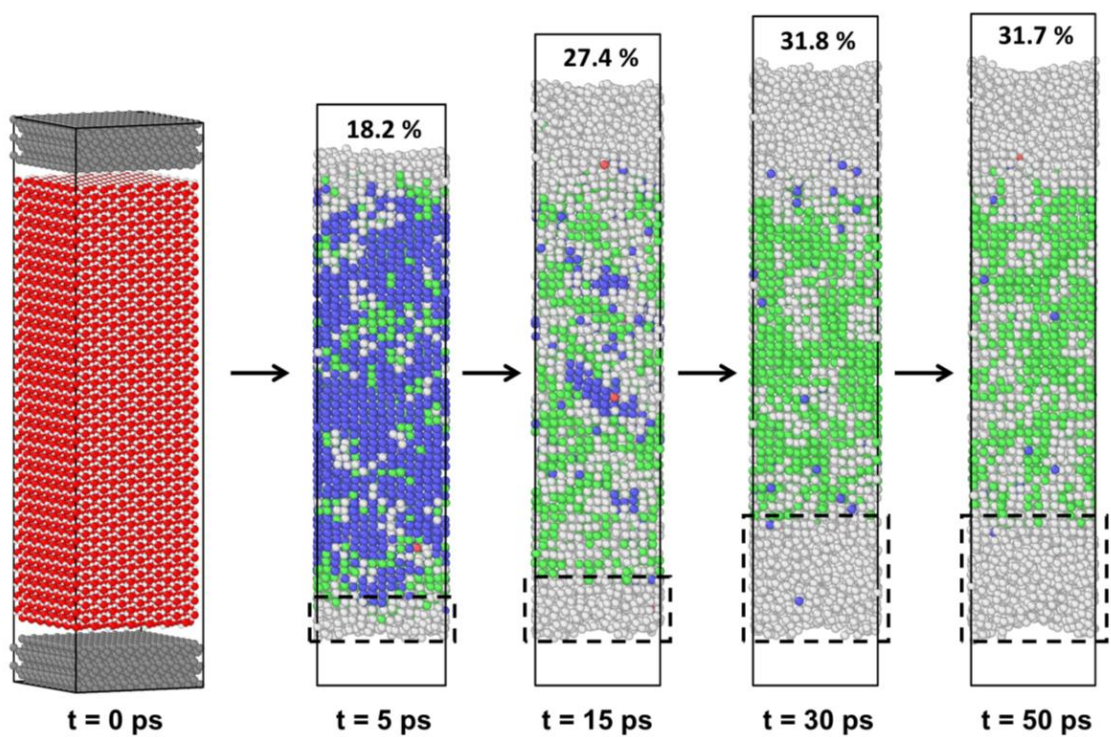

**FIG. S15.** Snapshots of 50 ps MD simulation for diamond/bcc-ice system at 20 GPa and 950 K. Oxygen atoms are color-coded as: green (fcc), blue (bcc), red (hcp) for ordered sublattices, and gray (disordered).

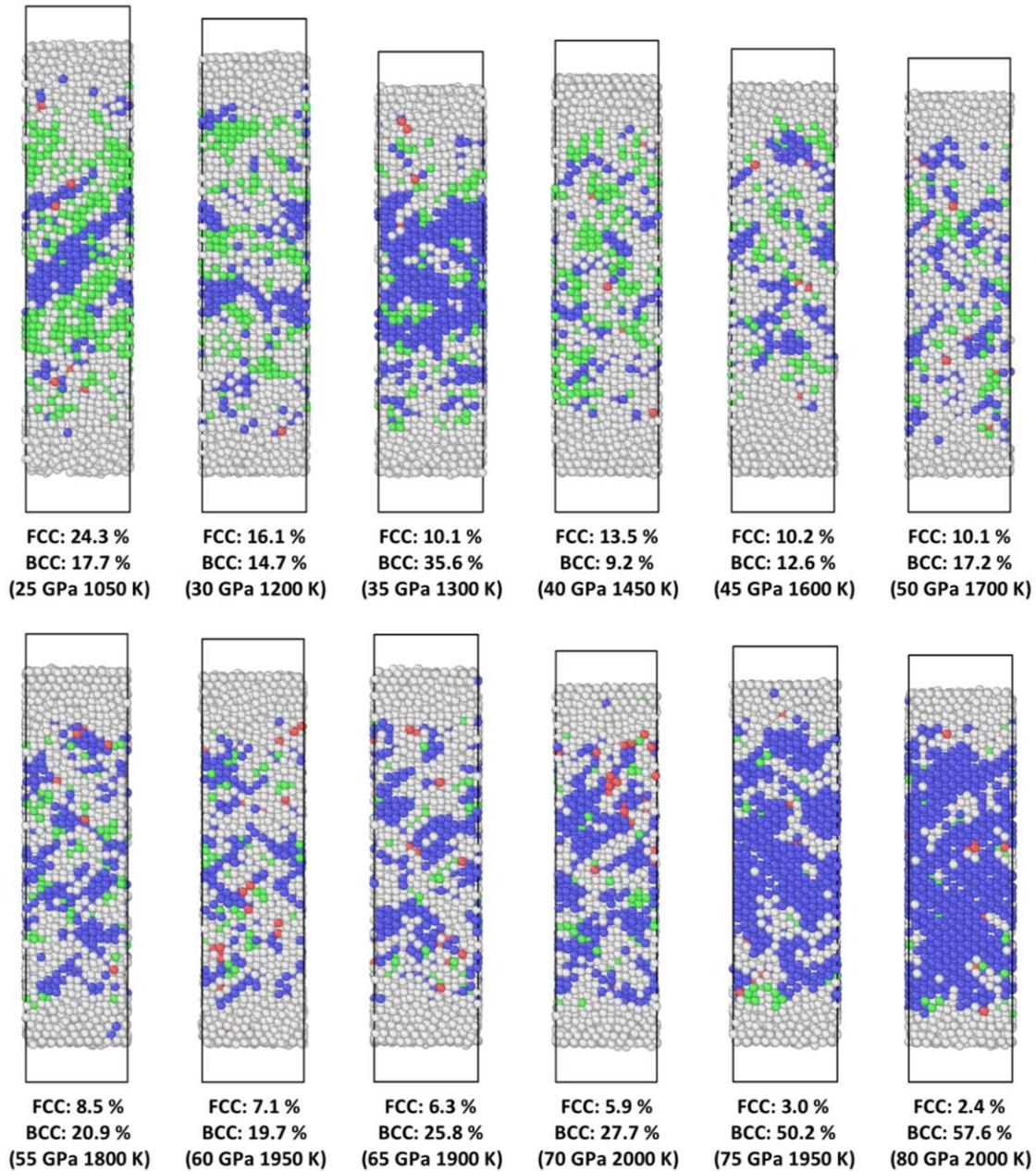

**FIG. S16.** The coexist phases of bcc and fcc ice observed in the other pressure ranges (from 25 GPa to 80 GPa). Oxygen atoms are color-coded as: green (fcc), blue (bcc), red (hcp) for ordered sublattices, and gray (disordered).

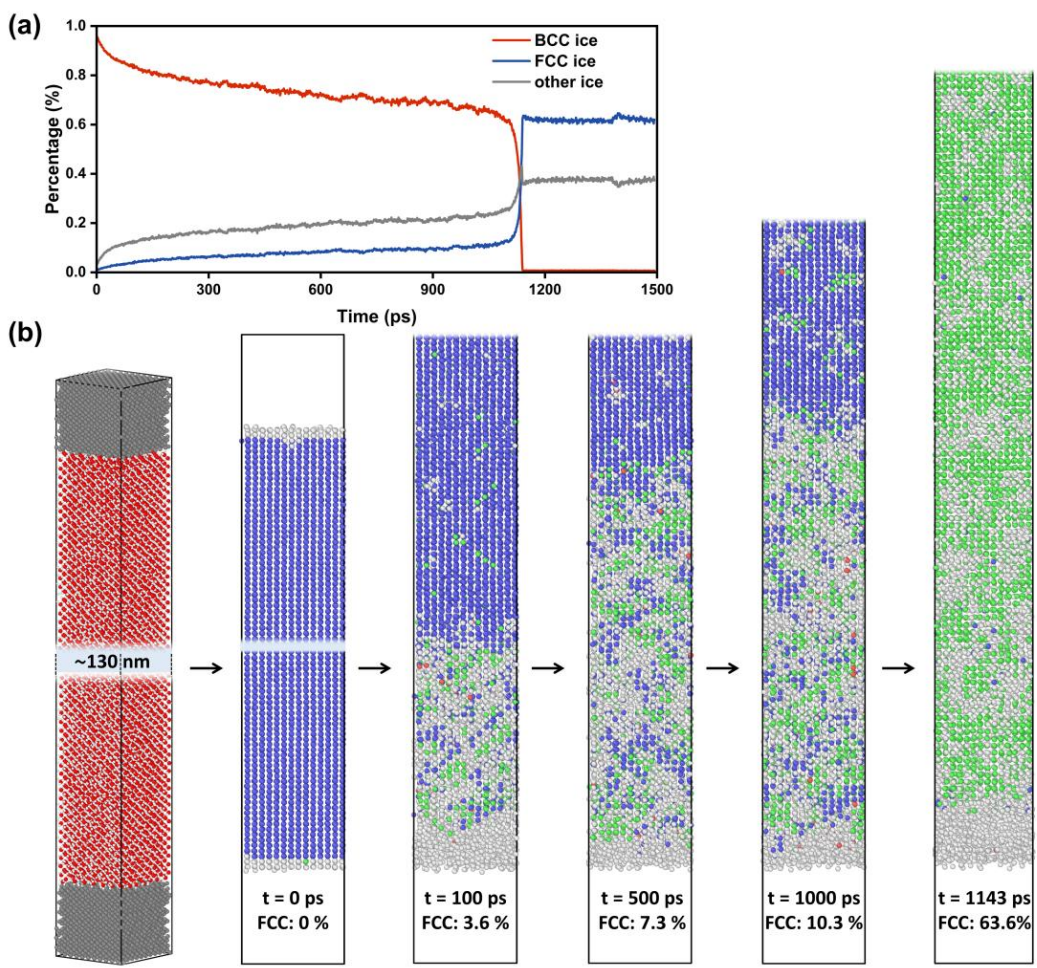

**FIG. S17.** (a) Time evolution of ice structure proportions in the extended system (233,346 atoms, 130 nm along the C-axis), and (b) corresponding structural snapshots. Oxygen atoms are color-coded as: green (fcc), blue (bcc), red (hcp) for ordered sublattices, and gray (disordered).

**Supplemental Reference:**

[1] Y. Zhang, H. Wang, W. Chen, J. Zeng, L. Zhang, H. Wang, and W. E, Computer Physics Communications **253**, 107206 (2020).

[2] H. Wang, L. Zhang, J. Han, and W. E, Computer Physics Communications **228**, 178 (2018).

[3] Z. Fan, W. Chen, V. Vierimaa, and A. Harju, Computer Physics Communications **218**, 10 (2017).

[4] K. Song *et al.*, Nature Communications **15**, 10208 (2024).

[5] G. Kresse and J. Furthmüller, Computational Materials Science **6**, 15 (1996).

[6] G. Kresse and D. Joubert, Physical Review B **59**, 1758 (1999).

[7] J. P. Perdew, K. Burke, and M. Ernzerhof, Physical Review Letters **77**, 3865 (1996).

[8] S. Grimme, J. Antony, S. Ehrlich, and H. Krieg, The Journal of Chemical Physics **132**, 154104 (2010).

[9] S. Grimme, S. Ehrlich, and L. Goerigk, Journal of Computational Chemistry **32**, 1456 (2011).

[10] T. D. Kühne *et al.*, Journal of Chemical Physics **152**, 194103 (2020).

[11] S. Plimpton, Journal of Computational Physics **117**, 1 (1995).

[12] S. Nosé, The Journal of Chemical Physics **81**, 511 (1984).

[13] M. Bernetti and G. Bussi, The Journal of Chemical Physics **153**, 114107 (2020).

[14] M. Ceriotti, M. Parrinello, T. E. Markland, and D. E. Manolopoulos, The Journal of Chemical Physics **133**, 124104 (2010).

[15] I. R. Craig and D. E. Manolopoulos, Journal of Chemical Physics **121**, 3368 (2004).

[16] N. Michaud-Agrawal, E. J. Denning, T. B. Woolf, and O. Beckstein, Journal of Computational Chemistry **32**, 2319 (2011).

[17] V. Wang, N. Xu, J.-C. Liu, G. Tang, and W.-T. Geng, Computer Physics Communications **267**, 108033 (2021).

[18] K. Momma and F. Izumi, Journal of Applied Crystallography **44**, 1272 (2011).

[19] A. Stukowski, Modelling and Simulation in Materials Science and Engineering **18**, 015012 (2010).